\documentclass[preprint,superscriptaddress,nofootinbib]{revtex4-1}
\usepackage{graphicx}
\usepackage{dcolumn}
\usepackage{bm}
\usepackage{amsmath}
\usepackage{amsfonts} 
\usepackage{latexsym}
\usepackage{bbm}
\usepackage{color}
\usepackage{amssymb}
\usepackage{comment}
\usepackage{amsthm}
\usepackage{epsf}
\usepackage{epsfig}
\usepackage{caption3}
\usepackage{appendix}
\usepackage{cases}
\usepackage{subfigure}
\usepackage{multirow}
\usepackage{float}

\usepackage{booktabs}
\usepackage{threeparttable}

\newcommand{\beq}{\begin {equation}}
\newcommand{\eeq}{\end   {equation}}
\newcommand{\bea}{\begin {eqnarray}}
\newcommand{\eea}{\end   {eqnarray}}
\newcommand{\baa}{\begin {array}   }
\newcommand{\eaa}{\end   {array}   }
\newcommand{\bit}{\begin {itemize} }
\newcommand{\eit}{\end   {itemize} }
\newcommand{\be }{\begin {equation}}
\newcommand{\ee }{\end   {equation}}
\newcommand{\nn }{\nonumber        }
\newcommand{\eepm}{e^{+}e^{-}}
\newcommand{\mm}{\mu^{+}\mu^{-}}

\newcommand{\leplep}{l^{+}l^{-}}
\newcommand{\lpm}{l^{\pm}}

\newcommand{\bbbar}{b\bar{b}}
\newcommand{\ccbar}{c\bar{c}}

\newcommand{\qqbar}{q\bar{q}}

\newcommand{\mz}{M_{Z}}

\newcommand{\mrecbb}{M^{\texttt{rec}}_{b\bar{b}}}
\newcommand{\mrecmm}{M^{\texttt{rec}}_{\mu^{+}\mu^{-}}}
\newcommand{\mmu}{M_{\mu^{+}\mu^{-}}}
\newcommand{\mrecee}{M^{\texttt{rec}}_{e^{+}e^{-}}}
\newcommand{\mee}{M_{e^{+}e^{-}}}
\newcommand{\ptee}{P^{T}_{e^{+}e^{-}}}
\newcommand{\ptll}{P^{T}_{l^{+}l^{-}}}
\newcommand{\mrecll}{M^{\texttt{rec}}_{l^{+}l^{-}}}
\newcommand{\mll}{M_{l^{+}l^{-}}}
\newcommand{\ptmm}{P^{T}_{\mu^{+}\mu^{-}}}
\newcommand{\mmb}{M_{b\bar{b}}}

\newcommand{\mh}{M_{h}}

\begin{document}

\title{Complementary probe of dark matter blind spots by lepton colliders and gravitational waves}

\author{Yan Wang}
\email[]{wangyan@imnu.edu.cn}
\affiliation{College of Physics and Electronic Information, Inner Mongolia Normal University, Hohhot 010022, China}
\affiliation{Inner Mongolia Key Laboratory for Physics and Chemistry of Functional Materials,
Inner Mongolia Normal University, Hohhot, 010022, China}
\author{Chong Sheng Li}
\email[]{csli@pku.edu.cn}
\affiliation{School of Physics and State Key Laboratory of Nuclear Physics and Technology, Peking University, Beijing 100871, China}
\affiliation{Center for High Energy Physics, Peking University, Beijing 100871, China}
\author{Fa Peng Huang}
\email[]{Corresponding author.  huangfp8@sysu.edu.cn}
\affiliation{MOE Key Laboratory of TianQin Mission, TianQin Research Center for Gravitational Physics $\&$ School of Physics and Astronomy, Frontiers Science Center for TianQin, CNSA Research Center for Gravitational Waves, Sun Yat-sen University (Zhuhai Campus), Zhuhai 519082, China}
\affiliation{Department of Physics and McDonnell Center for the Space Sciences, Washington University, St.Louis, MO 63130, USA}
\begin{abstract}
We study how to unravel the dark matter blind spots by phase transition gravitational waves in synergy
with collider signatures at electroweak one-loop level taking the inert doublet model as an example.
We perform a detailed Monte Carlo study at the future lepton colliders in the favored parameter space,
which is consistent with current dark matter experiments and collider constraints. Our studies demonstrate
that the Circular Electron Positron Collider and other future lepton colliders have the potential to explore
the dark matter blind spots.
\end{abstract}

\maketitle

\section{Introduction}

In recent years, there is a growing number of cosmological and astrophysical evidence on the existence of the
mysterious dark matter (DM) including the galaxy rotation
curve, the precise cosmic microwave background spectrum,
the bullet cluster collision, the gravitational lensing effects,
and so on~\cite{Bertone:2016nfn}. However, the absence of DM signals at the
DM direct search and LHC has almost pushed DM
parameter space to the blind spots, where the coupling
between DM and the standard model (SM) particles is too
small to be detected directly in the DM detectors. This
situation may point us towards some new approaches to
explore these DM parameter spaces, such as the future
gravitational wave (GW) experiments and the future lepton
colliders. After the discovery of GW by LIGO, GW
becomes a novel and realistic approach to understand
and explore the fundamental physics, including the mysterious DM. 
Meanwhile, the proposed future lepton colliders may also help to unravel the DM nature due to their
clean backgrounds and high sensitivity.

In this work, we revise the well-studied inert doublet
model (IDM)~\cite{Barbieri:2006dq}, which can provide natural DM candidates~\cite{Barbieri:2006dq,Honorez:2010re}. The current DM direct search has constrained
the Higgs-DM coupling to be very small. The DM direct
search might be difficult to observe the possible DM
signals. However, the inert scalars including the DM could
trigger a strong first-order phase transition (SFOPT) and
produce the phase transition GWs. Meanwhile, they could
modify the Higgs-Z boson coupling and the triple Higgs
coupling through loop effects. These modifications could
be exploited by the precise measurements of the process
$e^{+}e^{-} \to hZ$ with its various decay channels at future lepton
colliders, such as Circular Electron Positron Collider (CEPC)~\cite{CEPCStudyGroup:2018ghi,An:2018dwb}, Future Circular Collider (FCC-ee)~\cite{Abada:2019zxq}, and International Linear Collider (ILC)~\cite{Fujii:2017vwa}. In this work, we
focus on the detailed Monte Carlo (MC) simulations of the
lepton collider signals up to one-loop level in complement
to the corresponding GW signals induced by this DM
model. The details of SFOPT and collider simulations are
given in the Appendixes A and B.

The work is organized as the following:
In section~\ref{md}, we review the IDM, the DM blind spots from various constraints and the condition of a SFOPT. The detailed discussions of the phase transition GW spectra are given in section~\ref{gws}.
Then we focus on the MC simulations of the signals at future lepton colliders at the one-loop level in section~\ref{cmc}.
Lastly, the conclusion is given in section~\ref{sum}.

\section{Dark matter and strong first-order phase transition in the inert doublet model}\label{md}
The well-studied IDM could provide a natural DM
candidate and improve the naturalness~\cite{Barbieri:2006dq,Honorez:2010re}. This model
could also produce a SFOPT~\cite{Chowdhury:2011ga}. The tree-level scalar
potential at zero temperature of the IDM can be written as
the following:
\begin{align}
V =&\mu^2_1|\Phi|^2+\mu^2_2|\eta|^2
			+\frac{1}{2}\lambda_1|\Phi|^4+\frac{1}{2}\lambda_2|\eta|^4  \nonumber\\
			&+\lambda_3|\Phi|^2|\eta|^2
			+\lambda_4|\Phi^\dagger\eta|^2
			+\frac{1}{2}\{ \lambda_5(\Phi^\dagger\eta)^2+h.c.\}\,\,, \label{eq:potential}
\end{align}
where $\Phi$ is the SM Higgs doublet and $\eta$ is the inert doublet.
The vacuum stability puts the conditions~\cite{Barbieri:2006dq,Honorez:2010re}
\begin{align}
  \lambda_1^{} > 0,\ \lambda_2^{} > 0,\ \sqrt{\lambda_1^{} + \lambda_2^{}} + \lambda_3^{} > 0,\ \lambda_3^{}+\lambda_4^{} \pm |\lambda_5^{}| > 0\,\,. \label{eq:vs}
\end{align}
At zero temperature, the two doublet scalar fields can be expanded as
\begin{equation}
	\Phi=
	\begin{pmatrix}
 		G^+\\
 		\frac{1}{\sqrt{2}}(h+v+i G^0)
	\end{pmatrix}
	,\  \eta=
	\begin{pmatrix}
 		H^+\\
 		\frac{1}{\sqrt{2}}(H+iA)
	\end{pmatrix},
\end{equation}
where SM Higgs boson $h$ has 125~GeV mass and the vacuum expectation value (VEV) $v=246$~GeV.
$G^+$ and $G^0$ are the Nambu-Goldstone bosons. 
At zero temperature, the scalar masses can be obtained as
\begin{align}
	m_h^2&=\lambda_1 v^2, \\
	m_H^2&=\mu_2^2+\frac{1}{2}(\lambda_3+\lambda_4+\lambda_5)v^2, \\
	m_A^2&=\mu_2^2+\frac{1}{2}(\lambda_3+\lambda_4-\lambda_5)v^2, \\
		m^2_{H^{\pm}}&=\mu_2^2+\frac{1}{2}\lambda_3 v^2 \,\,.
\end{align}
These new inert scalars could contribute to the  modification of the $T$ parameter $\Delta T$, which could be approximated as 
\begin{align}
 \Delta T \simeq \frac{1}{6\pi e^2 v^2}(m_{H^{\pm}}-m_H)(m_{H^{\pm}}-m_A)\,\,.
\end{align}
If $m_A^2=m^2_{H^\pm}$ or $m_H^2=m^2_{H^\pm}$, $\Delta T \simeq 0$.
A simple and natural way to avoid large $T$ parameter deviation $\Delta T$ is to assume 
$m_A^2=m^2_{H^\pm}$. To satisfy this condition, one assumes
\begin{equation}
\lambda_4=\lambda_5<0,~~\lambda_3>0 \,\,,
\end{equation}
which would be consistent with all the constraints from
electroweak precise measurements, DM direct searches
and the collider data. The T parameter constraint does not
require the signs of these couplings. Thus, the choices of
these signs are just for simplicity and for the constraints
from the DM direct searches and the collider data, as in
Ref.~\cite{Chowdhury:2011ga}. Therefore, we have degenerated pseudoscalar and
charged scalar masses

\begin{equation}
m_A^2=m^2_{H^\pm}=\mu_2^2+\frac{1}{2}\lambda_3 v^2   \,\,.
\end{equation}
Under the above assumptions,  the particle $H$ ($m_H^2=\mu_2^2+\lambda_{L}v^2$) is the lightest particle and can be the natural DM candidate~\cite{Barbieri:2006dq,Honorez:2010re}.
And the DM-Higgs boson coupling is defined as $\lambda_{L}=(\lambda_3+\lambda_4+\lambda_5)/2$.
The loop correction does not change our results since $\Delta T$ is very small by assuming $m_A^2=m^2_{H^\pm}$. And DM constraint has a slight modification after including the loop correction.

However, the DM direct search has put strong constraints
on this DM-Higgs coupling for different DM masses.
For example, the XENON1T data have pushed the
DM-nucleon spin-independent elastic scatter cross section
up to 
 $\sigma_{\text{SI}}=4.1 \times 10^{-47}~\text{cm}^2$ for about 30~GeV DM mass at $90\%$ confidence level~\cite{Aprile:2018dbl}. These constraints almost reach the blind spots of the IDM,
 which means the DM-Higgs coupling $\lambda_L$ should be extremely 
 small. The favored
channel is the Higgs funnel region, where the DM mass
is about half of the Higgs boson mass ($m_H  \simeq m_h/2$). For this Higgs funnel region, we can estimate the cross
section as
 \begin{eqnarray}
	\sigma_{\rm SI} \simeq  \frac{\lambda_L^2f_N^2}{\pi} \left(\frac{m_N^2}{m_{H} m_h^2}\right)^2
\end{eqnarray}
 with $f_N \simeq 0.3$. Here, we first do some simple estimations
using the above equation to get the constraint of DM-Higgs
coupling from the DM direct search. Here and after, we use
$\sc{micrOMEGAs}$~\cite{Barducci:2016pcb} to do the precise calculations. We show
the constraint below

 \begin{equation}
\lambda_L \lesssim 0.003 \,\,.
 \end{equation}
These blind spots are difficult for future direct observation
of DM signal at DM direct search experiments. In this
work, we study how to use future lepton colliders in synergy with GW to explore the DM blind spots. The
corresponding DM relic abundance for the blind spots
of the Higgs funnel region should satisfy the Plank 2018
result~\cite{Aghanim:2018eyx}:
\begin{equation}
\Omega_{\text{DM}} h^2 =  0.11933 \pm 0.00091\,\,.
\label{dm_relic}
\end{equation}
Since the DM mass is about half of the Higgs mass,
the dominant DM annihilation process is the Higgs-mediated $s$ channel $HH\to h \to W^{\pm*}W^{\mp}$ with off-shell $W$ boson, namely, the ratio of the channel's contribution is about $52\%$. The second important channel is the $HH\to h \to b \bar{b}$ with the contribution about $32\%$.
We use the $\sc{micrOMEGAs}$~\cite{Barducci:2016pcb} to precisely calculate the DM relic abundance including the important resonant effects.

Besides the still allowed DM candidate in the blind
spots, the IDM in the blind spots could also trigger a
SFOPT, which can further produce phase transition GW
signals, and have a possibility to explain the electroweak
baryogenesis. 
When $\lambda_3$, $\lambda_4$, $\lambda_5$ are $\mathcal{O}(1)$, a SFOPT can be triggered~\cite{Chowdhury:2011ga,Borah:2012pu,
Gil:2012ya,Cline:2013bln,AbdusSalam:2013eya,
Blinov:2015vma,Cao:2017oez,Huang:2017rzf,
Laine:2017hdk,Senaha:2018xek,Huang:2019riv,
Kainulainen:2019kyp}.
The subtle point is the cancellation
between the three couplings, which can make the DM-Higgs coupling very small to satisfy the DM direct search.
We show the detailed discussions of the phase transition in
Appendix A.

Numerically, we use the package $\sc{micrOMEGAs}$~\cite{Barducci:2016pcb} to consider all the precise constraints from DM relic abundance $\Omega_{\text{DM}}h^2$, DM direct search $\sigma_{\text{SI}}$, collider constraints~\cite{Belyaev:2016lok}, and use 
$\sc{CosmoTransitions}$~\cite{Wainwright:2011kj} to calculate the phase transition dynamics.
Taking all the above discussions into consideration, we choose the following benchmark point
$m_h=125$ GeV
$m_A=m_{H^\pm}=300$ GeV,
$m_H=62.66$ GeV,
$\mu_2=61.69$ GeV, which corresponds to $\lambda_L=(m_H^2-\mu_2^2)/v^2=0.002$\footnote{For this benchmark point set, $\lambda_4=\lambda_5\approx -1.4$, $\lambda_3\approx 2.8$.
Substituting these values in the unitarity conditions given in Appendix A of Ref.~\cite{Branco:2011iw}, we find the unitarity bound is satisfied.
}.
This benchmark point set can
explain the whole DM and satisfy the DM direct search.
Taking this set of benchmark points, the relic density,
DM direct search, collider constraints and a SFOPT can be
satisfied simultaneously.

\section{Gravitational wave spectra}\label{gws}
There are three well-known sources to produce phase
transition GWs during a SFOPT, namely, sound wave,
turbulence and bubble wall collisions. For most particle
physics models beyond the SM, the dominant source is the
sound wave mechanism, which usually produces more
significant and long-lasting signal~\cite{Hindmarsh:2013xza,Hindmarsh:2015qta,Hindmarsh:2017gnf} compared
to turbulence and bubble wall collisions. To obtain the
GW spectra, we need to calculate the phase transition
dynamics, which is quantified by several phase transition
parameters. We can calculate these parameters from the 
finite-temperature effective potential $V_{\rm eff}$, which is given in the appendix A. The first parameter is the phase transition strength parameter $\alpha$. There are several different definitions of $\alpha$, and we use the conventional definition below
\begin{equation}\label{convenalpha}
\alpha = \frac{\Delta V_{\rm eff} - T\frac{\partial\Delta V_{\rm eff}}{\partial T}}{\rho_R} \,\,,
\end{equation}
where $\rho_R= \pi^2g_{\rm eff} T_+^4/30$. $T_+$ should be the temperature of the plasma surrounding the bubbles where GWs have been produced.
It is important to choose the correct $T_+$~\cite{Wang:2020jrd}, which is usually chosen as the nucleation temperature $T_n$ (At $T_n$, one bubble is nucleated in one Hubble radius.) or the percolation temperature $T_p$ (At $T_p$,
about $34\%$ of false vacuum has been converted to true vacuum and a large numbers of bubbles have collided and percolated.).
The phase transition strength parameters calculated at the nucleation temperature $T_n$
and the percolation temperature $T_p$ are denoted by $\alpha_n$ and $\alpha_p$, respectively. 
The second parameter is the mean bubble separation $R_{*}$, which is given by
\begin{equation}
R_* = n_b^{-1/3}\,\,, \label{HR}
\end{equation}
where $n_b$ is the bubble number density~\cite{Turner:1992tz}.

From the recent numerical simulations~\cite{Hindmarsh:2013xza,Hindmarsh:2015qta,Hindmarsh:2017gnf}, the simulated GW spectrum from the sound wave can be written as
\begin{equation}
h^2\Omega_{\rm sw}(f) \simeq 1.64\times10^{-6}(H_*\tau_{\rm sw})(H_*R_*)K^2\left(\frac{100}{g_*}\right)^{1/3}(f/f_{\rm sw})^3\left(\frac{7}{4 + 3(f/f_{\rm sw})^2}\right)^{7/2},\label{swf}
\end{equation}
with the peak frequency
\begin{equation}
f_{\rm sw} \simeq 2.6 \times10^{-5}~\text{Hz}\frac{1}{H_*R_*}\left(\frac{T_*}{100~ \rm GeV}\right)\left(\frac{g_*}{100}\right)^{1/6} \,\,.
\end{equation}
$\tau_{\rm sw}$ is the sound wave duration time,
\begin{equation}
\tau_{\rm sw} = \min\left[\frac{1}{H_*}, \frac{R_*}{\overline{U}_f}\right]   \,\,,
\end{equation}
and the kinetic energy fraction
\begin{equation}
K=\frac{\kappa_v\alpha}{1 + \alpha}    \,\,.
\end{equation}
$H_*$ is the Hubble parameter at $T_*$. The efficiency parameter $\kappa_v$ is the fraction of vacuum energy converted into the fluid bulk kinetic energy.
The root-mean-square fluid velocity $\overline{U}_f^2 $ is approximated as \cite{Hindmarsh:2017gnf, Caprini:2019egz, Ellis:2019oqb}
\begin{equation}
\overline{U}_f^2 \approx \frac{3}{4}K \,\,.
\end{equation}
The duration time $\tau_{\rm sw}$ determines 
whether the sound wave spectrum is suppressed or not.
Qualitatively, for $ H_*  \tau_{\rm sw}<1$, the GW spectrum is suppressed by a factor of $H_*R_*/\overline{U}_f$, namely, $\Omega_{\rm sw} \propto K^{3/2}$.
In the opposite direction, there is no suppression, and
the GW spectrum scales as
$\Omega_{\rm sw} \propto K^2$.

Many models predict the suppressed sound wave spectrum, and hence the contributions from turbulence and
bubble collisions might not be negligible. The GW spectrum from turbulence is still controversial~\cite{Kosowsky:2001xp,Gogoberidze:2007an,Niksa:2018ofa,Caprini:2019egz} and
we use the following formula as an estimation~\cite{Caprini:2009yp,Caprini:2015zlo}:
\begin{equation}
h^2\Omega_{\rm turb}(f) \simeq 1.14\times10^{-4}H_*R_*\left(\frac{\kappa_{\rm turb}\alpha}{1 + \alpha}\right)^{3/2}\left(\frac{100}{g_*}\right)^{1/3}\frac{(f/f_{\rm turb})^3}{(1 + f/f_{\rm turb})^{11/3}(1 + 8\pi f/H_*)} \,\,.
\end{equation}
The efficiency factor 
 $\kappa_{\rm turb}$ is given by the  recent simulations~\cite{Hindmarsh:2015qta}.
The Hubble rate at $T_*$ is given by
\begin{equation}
H_* = 1.65 \times 10^{-5}~\text{Hz}\left(\frac{T_*}{100~ \rm GeV}\right)\left(\frac{g_*}{100}\right)^{1/6} \,\,.
\end{equation}
Thus, we can obtain the peak frequency of turbulence $f_{\rm turb}$
\begin{equation}
f_{\rm turb} \simeq 7.91\times10^{-5}~\text{Hz}\frac{1}{H_*R_*}\left(\frac{T_*}{100 ~\rm GeV}\right)\left(\frac{g_*}{100}\right)^{1/6}.
\end{equation}

To obtain more reliable GW spectra, we need to first
know the bubble wall velocity and energy budget which are
explicitly model dependent. The GW spectra strongly
depend on the bubble wall velocity. Most of the previous
studies on the GW spectra in a given new physics model
just take the bubble wall velocity as an input parameter.
Explicitly, the bubble wall velocity is determined by the
friction force of thermal plasma acting on the bubble wall.
The friction force is further determined by the deviation of
massive particle populations from the thermal equilibrium.
Here, we estimate a more realistic bubble wall velocity as
$v_b=0.3$ based on the
Refs.~\cite{Moore:1995si,Wang:2020zlf}, 
where the friction force is
similar to the SM case. The precise calculations of the
bubble wall velocity for a given new physics model are
complicated, which is beyond the scope of this paper. We
notice that this model is similar to the model discussed in
Refs.~\cite{Moore:1995si,Wang:2020zlf}, and hence choose the approximated values
as in these references. This bubble wall velocity is smaller
than the sound speed
$v_s=\sqrt{3}/3$, and thus this case
belongs to the deflagration mode, which can be further
used to successfully explain the electroweak baryogenesis.
For the energy budget, the model-independent formula is
used in most of the previous studies. The model-dependent 
studies find that there are modifications of the energy
budget considering more realistic sound speed in the
broken phase and symmetric phase during a SFOPT~\cite{Giese:2020znk,Wang:2020nzm}. However, the phase transition strength is weak
and the corresponding correction of the energy budget
is not significant in this work~\cite{Giese:2020znk,Wang:2020nzm}. We can still use
the model-independent energy budget formula as an
estimation.

We could first give some qualitative discussions on our
predictions of the GW spectra. In our previous work~\cite{Wang:2020jrd},
we classified the SFOPT into four cases. This model
belongs to the weakest type, namely, the slight supercooling, which corresponds to $\alpha_p \leq 0.1$.
In this case, $\alpha_n$ can be a good approximation to $\alpha_p$ since $\alpha_p-\alpha_n \ll 0.1$.
For slight supercooling, the GW signal is too weak and difficult
to be detected by the Laser Interferometer Space Antenna
(LISA)~\cite{Caprini:2019egz}. The signal may be within the sensitivity
of Ultimate-Decihertz Interferometer Gravitational wave
Observatory (U-DECIGO)~\cite{Kudoh:2005as}, and big bang observer
(BBO)~\cite{Corbin:2005ny}.

\begin{figure}[htbp]
\begin{center}
\includegraphics[width=14cm]{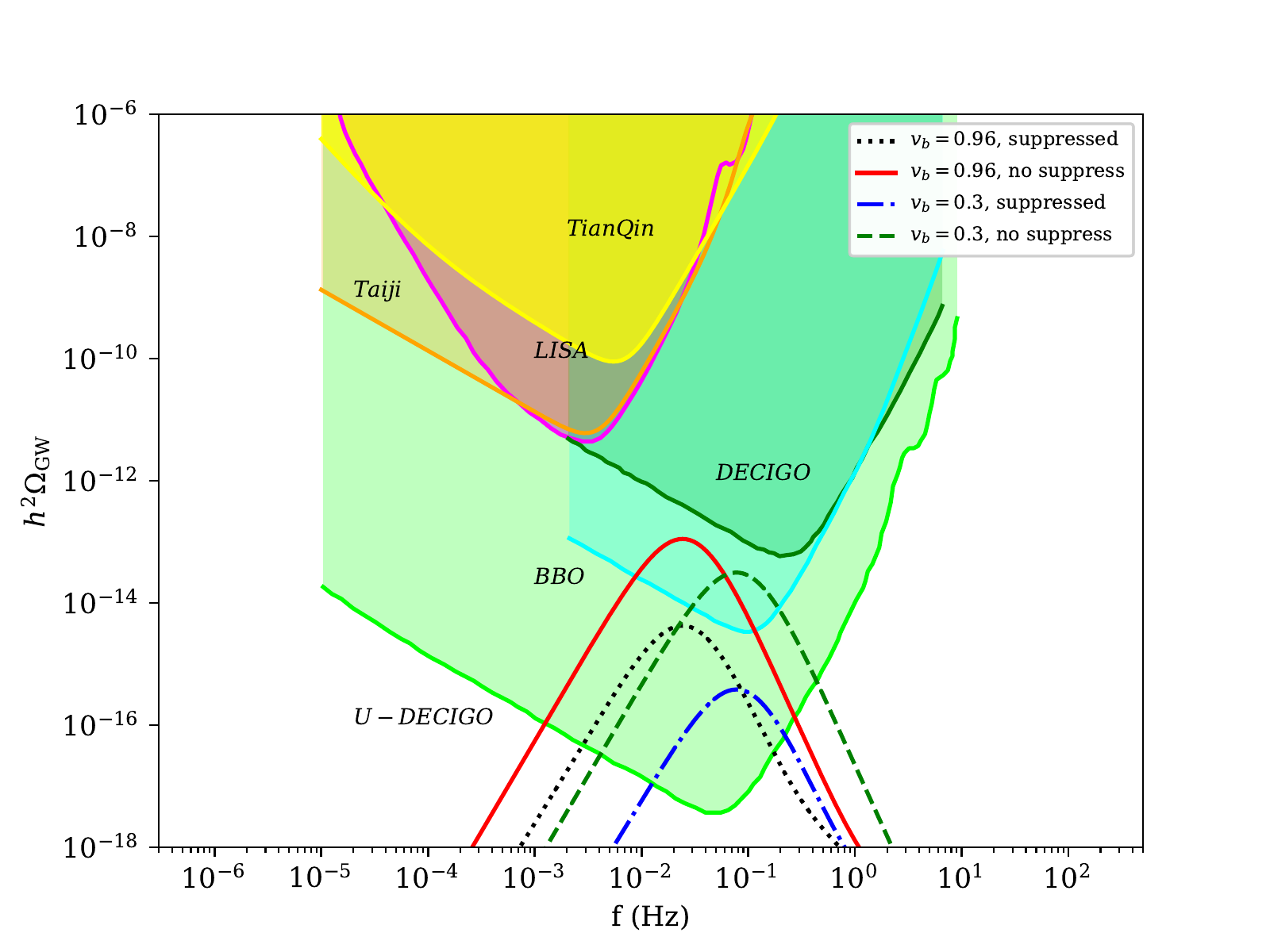}
\caption{The GW spectra for IDM. The colored regions represent the expected sensitivities of the future GW experiments. Different lines depict different GW spectra for different bubble wall velocities and sound wave duration.} 
\label{gw}
\end{center}
\end{figure}

Combining the above discussions, we show the GW
spectra from the three sources in Fig.~\ref{gw} for the benchmark
points. The colored regions represent the expected sensitivity for the future GW experiments, LISA~\cite{Audley:2017drz,Caprini:2015zlo,Caprini:2019egz,LISA:documents},
TianQin~\cite{Luo:2015ght,Hu:2018yqb,Mei:2020lrl}, Taiji~\cite{Hu:2017mde,Guo:2018npi}, DECIGO~\cite{Seto:2001qf,Kawamura:2011zz},
U-DECIGO~\cite{Kudoh:2005as}, and BBO~\cite{Corbin:2005ny}. The red line represents
the GW spectra for the bubble wall velocity $v_b=0.96$ without suppression, while  the black dotted line depicts the GW spectra 
for the bubble wall velocity $v_b=0.96$ with suppression.
The blue dash-dotted line and the green dashed line corresponds to the
GW spectra with and without suppression for the wall velocity $v_b=0.3$, respectively.
It is obvious that the bubble wall velocity and suppression effects are significant. For our benchmark points in the IDM, our estimation favors the wall velocity $v_b=0.3$ with suppression effect.
Since this type belongs to the slight supercooling case, the GW spectra are too weak to be detected by LISA, Taiji, TianQin, and DECIGO.
However, they can reach the sensitivity of U-DECIGO or BBO.

After having the GW spectra of the signal, the detectability of the GW signal needs to be quantified by defining
the conventional signal-to-noise ratio (SNR):
\begin{equation}
  \text{SNR} = \sqrt{\mathcal{T}_\text{obs} \int_{f_\text{min}}^{f_\text{max}}
    \mathrm{d}f \left[ \frac{h^2 \Omega_\text{GW} (f) }{ h^2
        \Omega_\text{det}(f)} \right]^2 } \, ,
  \label{snr}
\end{equation}
where $\mathcal{T}_\text{obs}$ is the total observation time and $h^2 \Omega_\text{det}(f)$ is the nominal sensitivity of a
given GW experiment configuration to cosmological sources.
We simply assume four-years mission duration time with a duty cycle of 75\% $\mathcal{T}_\text{obs}$, and take  $\mathcal{T}_\text{obs}\simeq 9.46\times 10^7\,$s, which is guaranteed by the LISA~\cite{LISA:documents}.
For the benchmark points with the wall velocity $v_b=0.3$ and the  suppression effect, the SNR is about nine. We can see that U-DECIGO is capable to detect the signals with enough observation time.

It is worth noticing that there are large theoretical
uncertainties in the predictions of the GW spectra. In the
above discussions, we clarify the dominant uncertainties
from model-dependent bubble wall velocity, definition of
the phase transition parameters, the suppression effects
in sound wave~\cite{Guo:2020grp}, model-dependent kinetic energy
fraction, and so on. In Fig.~\ref{gw}, we choose more conservative estimations in our calculations. Considering the large
uncertainties and taking progressive estimations, the GW
signal could be within the sensitivity of DECIGO and the
marginal region of LISA. In a recent study~\cite{Croon:2020cgk}, the three-dimensional approach could significantly reduce the uncertainties. We leave the three-dimension study for this IDM in
our future work.

\section{Precise predictions at future lepton colliders}\label{cmc}
The SM $hZ$ leading-order cross section ($e^+  e^- \to h Z \to \mm h$) at 240 GeV CEPC is 6.77 fb calculated by $\sc{Whizard~1.95}$~\cite{Kilian:2007gr}. 
At the lepton collider, the Higgs-strahlung process offers an unique opportunity for a model-independent precise measurement of the $hZZ$ coupling strength.
At the CEPC~\cite{CEPCStudyGroup:2018ghi,An:2018dwb} with an integrated luminosity of $5.6~\mathrm{ab}^{-1}$, the precision of $\sigma_{hZ}$ could achieve about $0.5\%$  with a ten-parameter fit to the CEPC and high luminosity LHC (HL-LHC) data~\cite{An:2018dwb}, which corresponds to the uncertainty of $hZZ$ coupling 0.25\%. The uncertainty could further reach 0.12\% with a seven-parameter effective field theory (EFT) fit~\cite{deBlas:2019wgy}.
At the ILC, the projected uncertainty of $hZZ$ coupling for the ILC EFT analysis could reach 0.18\% when combining HL-LHC, 250 GeV ILC  and 500 GeV ILC data~\cite{Bambade:2019fyw}. At the FCC-ee, combining HL-LHC, 240 GeV and 365 GeV FCC-ee data, the $hZZ$ uncertainty also reach 0.16\%. The above measurements are all based on the recoil mass technique to give model-independent constraints, where the $Z$ boson decays to $\eepm$, $\mm$ or $\qqbar$ and the Higgs boson decay final states do not need to be considered. In a specific model, when the Higgs decay mode could be determined, the $hZZ$ coupling could be measured more precisely. 

Due to the loop effects of the new particles, the $hZZ$ coupling strength is modified in the IDM. The one-loop electroweak corrections of the $hZZ$ vertex in the IDM are calculated in \cite{Arhrib:2015hoa, Kanemura:2016sos}. In this study, we adopt the one-loop electroweak corrections in the Ref.~\cite{Kanemura:2016sos}. After considering the one-loop electroweak radiative effects, 
the Lorentz structures of the $hZZ$ coupling become~\cite{Barklow:2017awn}: 
\begin{equation}
    \mathcal{L}_{hZZ} = M^{2}_{Z}(\frac{1}{v}+ \frac{a_Z}{2\Lambda})Z_{\mu}Z^{\nu}h +  \frac{b_Z}{2\Lambda} Z_{\mu\nu}Z^{\mu\nu}h + \frac{\tilde{b}_Z}{2\Lambda} {Z}_{\mu\nu}\tilde{Z}^{\mu\nu}h\,\,,
\end{equation}
where $Z_{\mu\nu} \equiv \partial_{\mu}Z_{\nu}-\partial_{\nu}Z_{\mu}$ and $\tilde{Z}_{\mu\nu} \equiv \frac{1}{2}\epsilon_{\mu\nu\rho\sigma}Z^{\rho\sigma}$.  
The detailed expressions of $a_{Z}$, $b_{Z}$, $\tilde{b}_{Z}$ can be found in Ref.~\cite{Kanemura:2016sos}.
The first term is similar to the SM and will affect the total cross section, while the second and the third term will affect final state angular distributions as well as the total cross sections. 

We integrate the next-leading-order (NLO) electroweak correction of the $hZZ$ vertex in the $\eepm \to \mm h$ process in the SM as well as in the IDM into the $\sc{Whizard}$ code.  The $hZ$ cross sections with the one-loop contributions to the $hZZ$ coupling at the different center-of-mass energy are listed in Table.~\ref{t:cross_section}. The deviation of the $\eepm \to \mm h$ NLO cross section between the IDM and the SM is defined as:
\begin{equation}
\Delta\sigma \equiv (\sigma_{IDM}^{NLO}-\sigma_{SM}^{NLO})/\sigma_{SM}^{NLO},
\end{equation}
where $\sigma_{IDM}^{NLO}$ and $\sigma_{SM}^{NLO}$ are the cross sections with the electroweak one-loop contributions of the $hZZ$ coupling in the IDM and in the SM. $\Delta\sigma$ is about $-0.2\%$ within our benchmark parameters at the 240 GeV. Although the deviation is slight, it still can be searched at the future electron-positron colliders with the model-independent measurements. It is worth noting that the deviation depends on the beam polarization,  $\Delta\sigma$ reaches the minimum for the pure left-hand electron and right-hand positron, where ILC could play an important role.

\begin{table}
 \begin{center}
 \begin{small}
     \caption{The electroweak one-loop cross sections for the $\eepm \to \mm h$ process in the SM model and the IDM model when the center-of-mass energies are 240 GeV and 250 GeV. The parameters are $m_h=125$ GeV, $m_A=m_{H^\pm}=300$ GeV, $m_H=62.66$ GeV, $\mu_2=61.69$ GeV.}\label{t:cross_section}
 \begin{tabular}{|c| c| c| c| c| c| c| c| c| c|}
 \hline
  $240$ GeV          & total $\sigma$ & $e^{-}_{L}e^{+}_{R}$ & $e^{-}_{R}e^{+}_{L}$ \\
 \hline
 SM  $\text{NLO}$ (fb)                         & 6.244           & 15.203                & 9.749  \\
 \hline
 IDM $\text{NLO}$  (fb)                          & 6.230           & 15.159                & 9.750 \\
  \hline
  $\Delta\sigma$    & -0.22\%           & -0.289\%                & 0\% \\
   \hline
  $250$ GeV         & &  &   \\
 \hline
 SM  $\text{NLO}$  (fb)                         & 6.615           & 16.158                & 10.376  \\
 \hline
 IDM  $\text{NLO}$  (fb)                        & 6.623           & 16.126                & 10.375 \\
   \hline
  $\Delta\sigma$    & -0.12\%           & -0.20\%                & 0\% \\
  \hline
 \end{tabular}
 \end{small}
 \end{center}
  \end{table}

Furthermore, considering that in the above future collider predictions, the common procedure is first to measure
the $hZ$ production cross section and the $hZZ$ coupling by
the model-independent measurement of the Higgs decay
final states with the recoil mass technique. Then, measure
the branching ratios of each Higgs decay channel. The
precision is sacrificed for the model independence.
However, since the deviation of the $hZ$ production cross
section in the IDM and SM is very small, it is hard to be
distinguished with the above model-independent measurements in those future colliders. In order to suppress the
background and increase the measurement significance, we
could directly measure the $\eepm \to Z h \to \leplep \bbbar$ process
to suppress the backgrounds with the explicit Higgs decay
channel $h\to \bbbar$, because in our scenario the Higgs boson
decay is considered the same as the SM Higgs. Then, the
result will be folded back to the $hZ$ cross section with
the SM $h\to \bbbar$ branching ration. In this case, a lot of
backgrounds in the model-independent analysis will be
exceedingly suppressed, such as two fermion production
($\eepm \to \leplep$), as well as four leptonic fermion production
$\eepm \to ZZ/WW \to llll (ll\nu\nu)$ and so on. Thus, the $hZZ$
coupling measurement resolution will increase, comparing
with the above predictions in the future colliders. Because
we cannot fully simulate the future collider MC analysis,
we will perform the fast simulation, analyze with the
explicit $h\to \bbbar$ model measurement and recoil mass
measurement, respectively. As a comparison, the model-independent measurement for the $hZZ$ coupling is listed in
the Appendix B. By comparing two results, we can
estimate the $hZZ$ constraints with the full simulation in
the future collider. It will be clear that the ability to search
the anomalous $hZZ$ coupling at the future Higgs factories
will be more greatly enhanced than the above $\Delta hZZ$
uncertainties with the model independent method.

We will perform the search for $\eepm \to Zh \to \leplep \bbbar$ in
the following section and state the recoil mass results in the
Appendix B to show the possible measurement accuracy.
By comparing two methods, we will understand how much
the $\Delta hZZ$ uncertainties are improved from the model
independent to the model dependent method. Thus, we
could estimate the possible uncertainties for the full
simulation at the future lepton colliders. The MC events
are simulated with the following features:
\begin{itemize}
	\item The signal events are generated by $\sc{Whizard}$ 1.95 with
unpolarized beams at the center-of-mass energy $\sqrt{s}=240$ GeV, where the one-loop electroweak
corrections to the $hZZ$ vertex in the IDM are coded
into the $\sc{Whizard}$.
	\item All other SM processes are considered as the
backgrounds, which are generated by $\sc{Whizard}$ 1.95
at the leading order. The details of the background
event generations at the CEPC can be found in
Ref.~\cite{Mo:2015mza}. According to the final-state fermion
number, the SM backgrounds are mainly classified
into three groups: two fermion case (2f) (include
Bhabha, $\eepm\to\mm/\tau\tau/qq$), four fermion case
(4f) (include $ZZ$/$WW$/single $Z/W$ production then $Z/W$ decay to fermions and so on), $hZ$
production and other $hZ$ decay channels except
the signal.
	\item The hadronization for the signal and background
events are accomplished by $\sc{Pythia6}$~\cite{Sjostrand:2006za}. The
bremsstrahlung and ISR effects are also considered
for both the signal and the background processes.
	\item All the event samples are then simulated with CEPC
detector configuration by using the default CEPC
detector card in the $\sc{Delphes}$-v3.4.2~\cite{deFavereau:2013fsa}. To cluster final
particles into jets, the anti-$k_{t}$ jet algorithm with jet parameter $\Delta R$=0.5 is applied with the $\sc{FastJet}$ package.
	\item $2\mu^{\pm}2b$ or $2e^{\pm}2b$ are required in the final states. The
b-tagging efficiency is 80\%, mistagging rate is 10\%
for c-qaurk jet and 0.1\% for light quark jets.
\end{itemize}

\subsection{Preselection}
At the first step, a pair of muon or electrons, whose
energies are larger than 5 GeV with different signs, are
selected. If there are more than two leptons in the event,
the lepton pair is selected by minimizing the following $\chi^2$-function:
\begin{equation}
\chi^{2} (\mll, \mrecll) = (\mll-\mz)^2 + (\mrecll-\mh)^2,
\end{equation} 
where $\mll$ is the invariant mass of the lepton pair and $\mrecll$ is the recoil mass of the lepton pair, which is defined as:
\begin{equation}
{\mrecll}^2 = (\sqrt{s}-E_{\leplep})^2-|\vec{p}_{\leplep}|^2.
\end{equation}
A further preselection cut is applied at this stage
for choosing the lepton pair: $\mll \in [50, 150]$ GeV, $\mrecll \in [50, 160]$ GeV.

After selecting the lepton pair, a photon is identified as
the bremsstrahlung or the final state radiation photon from a
lepton. If the polar angle of the photon with respect to the
lepton $\theta_{l^{\pm}-\gamma}$ is larger than $0.99$, the four-momentum of the
photon is combined to the lepton.

For the jets, we also require that two b-jets are tagged
with the leading jet energy larger than 20 GeV and the nextleading jet energy larger than 5 GeV. If there are more than
two b-jets, the $\chi^2$-function of the $\mmb$ and $\mrecbb$ is also used:

\begin{equation}
\chi^{2} (\mmb, \mrecbb) = (\mmb-\mh)^2 + (\mrecbb-\mz)^2,
\end{equation} 
where $\mmb$ is the invariant mass of the b-jet pair and $\mrecbb$ is the corresponding recoil mass.

\subsection{The multivariate analysis method}
\begin{figure}[ht]
      \begin{center}
         \includegraphics[height=8cm]{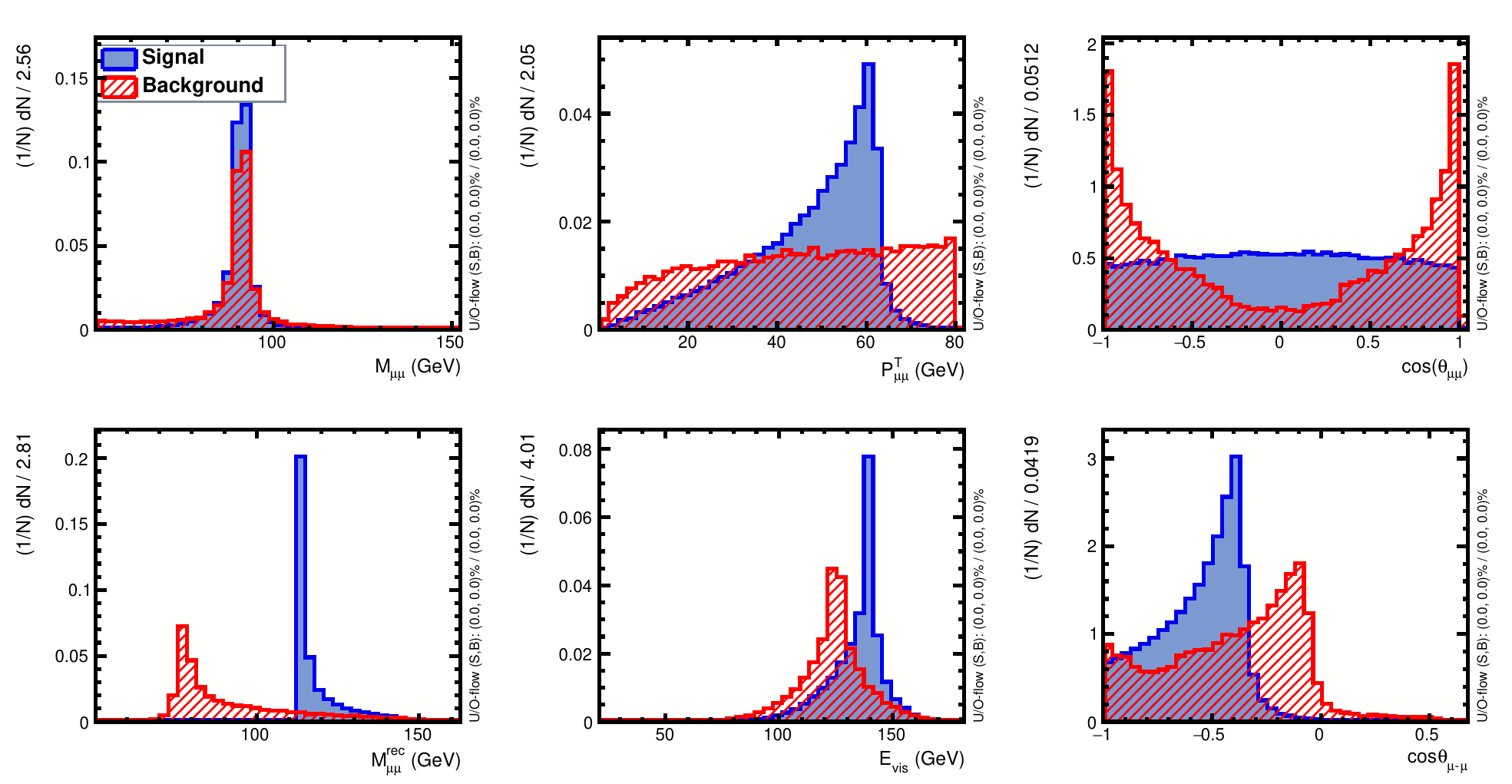}	
      \end{center}
      \begin{center}
         \includegraphics[height=8cm]{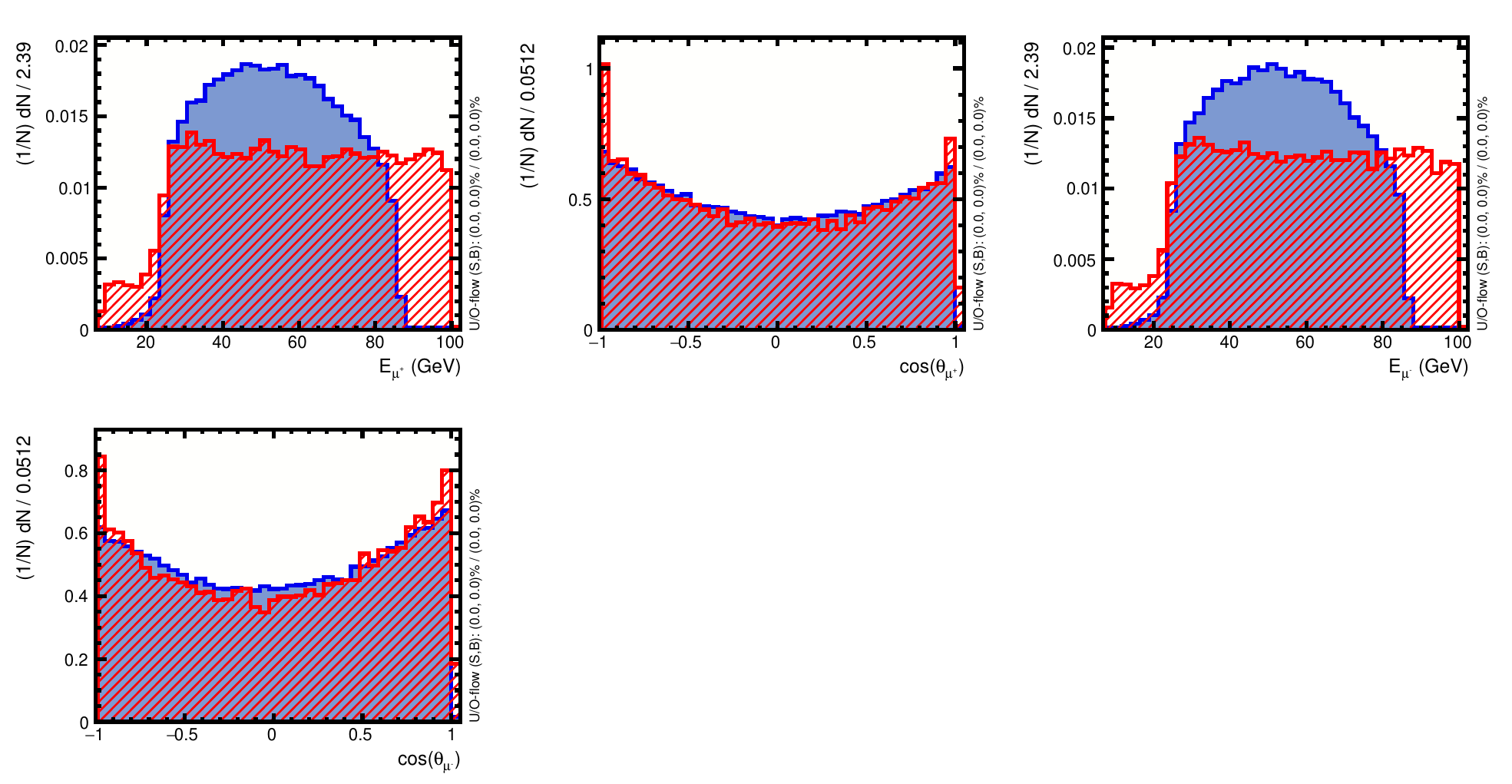}	
      \end{center}
    \caption{The input variables distributions for the lepton-related MVA. The variables are $\mmu$, $\ptmm$, $\textrm{cos}\theta_{\mm}$, $\mrecmm$, $E_{\text{vis}}$,  $\textrm{cos}\theta_{\mu^{+}-\mu^{-}}$,  $E_{\mu^{+}}$,  $\textrm{cos}\theta_{\mu^{+}}$,  $E_{\mu^{-}}$, $\textrm{cos}\theta_{\mu^{-}}$, respectively.}\label{f:MVA_mu_input}
\end{figure}

Two multivariate analysis (MVA) methods based on the
gradient boosted decision tree (BDTG) method~\cite{QUINLAN1987221}, which
is included in TMVA package~\cite{BRUN199781}, are used to improve the
sensitivity. The first BDTG is trained for the lepton-related
variables ($\text{MVA}_{\mu}$) with the signal and all possible background processes to wipe up reducible backgrounds.
$\text{MVA}_{\mu}$ is trained using the following ten input variables,
where the related input variable distributions can be found
in Fig.~\ref{f:MVA_mu_input}:

\begin{itemize}
\item the invariant mass of the lepton pair $\mll$, which
should be close to the $Z$ boson mass;
\item  the transverse momentum of the lepton pair $\ptll$
(for the signal, it should peak at about 60 GeV---in
contrast, for the background, it is rather flat and
widely distributed);
\item  the polar angle of the lepton pair cos $\textrm{cos}\theta_{\leplep}$ (the signal
$Zh$ events are the typical 2-to-2 production, while
two fermion events will prefer the beam region);
\item  the recoil mass of the lepton pair $\mrecll$ (for the signal,
it is close to the Higgs boson mass, while it will be
close to the $Z$ boson mass in the main backgrounds);
\item  the visible energy $E_{\text{vis}}$, which is defined as the sum
of the energies of all visible final states;
\item  the opening angle between the two leptons $\textrm{cos}\theta_{l^{+}-l^{-}}$;
\item  the lepton energies $E_{\lpm}$;
\item  the polar angle of each lepton $\textrm{cos}\theta_{l^{+}}$,  $\textrm{cos}\theta_{l^{-}}$.
\end{itemize}

\begin{figure}[ht]
      \begin{center}
         \includegraphics[height=8cm]{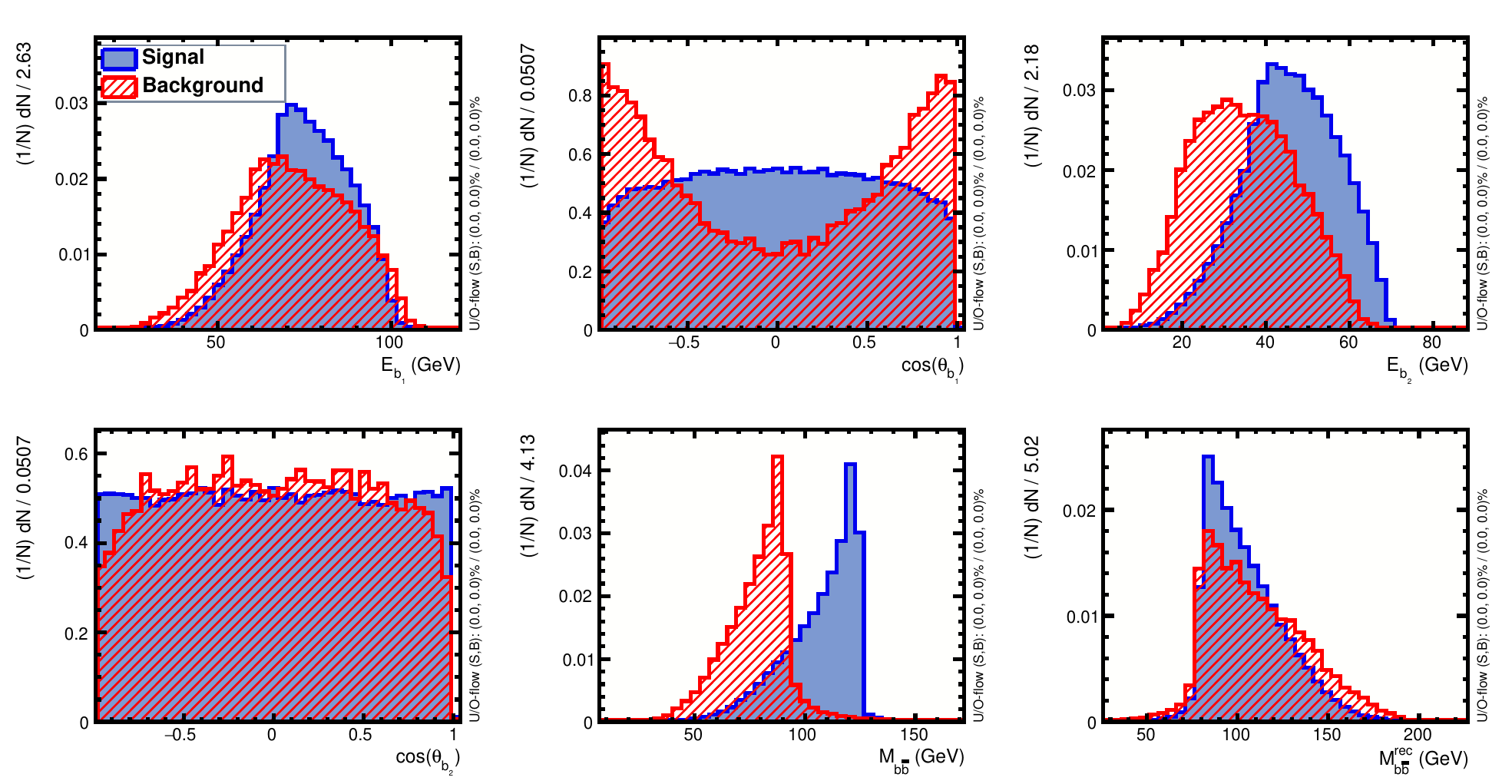}	
      \end{center}
    \caption{The input observable distributions for the jet-related MVA. The variables are  $E_{b_{1}}$, $\textrm{cos}\theta_{b_{1}}$, $E_{b_{2}}$, $\textrm{cos}\theta_{b_{2}}$, $\mmb$, $\mrecbb$, respectively.}, \label{f:MVA_jet_input}
\end{figure}

Since most of the reducible background will be discarded
with the lepton-related MVA cut and other kinematic cuts, the
second BDTG is only trained for the jet-related variables
($\text{MVA}_{j}$), which will be less noise disturbance. The jet-related
MVA will only train with the signal and $ZZ\to \leplep \bbbar$
process events, which is the main irreducible background.
The jet-related MVA is trained using the following six input
variables, where the related input variable distributions can
be found in Fig.~\ref{f:MVA_jet_input}:
\begin{itemize}
\item the energy of each b-jet $E_{b}$;
\item  the polar angle of each $\textrm{cos}\theta_{b}$;
\item  the invariant mass of the $\bbbar$, $\mmb$, which should be
close to the $h$ boson mass for the signal and be in
turn closed to the $Z$ boson mass for the background;
\item  the recoil mass of the lepton pair $\mrecbb$, which are
both close to the $Z$ boson mass for the signal and
background
\end{itemize}

\begin{figure}[ht]
\begin{minipage}{0.47\textwidth}
      \begin{center} 
         \includegraphics[height=5cm]{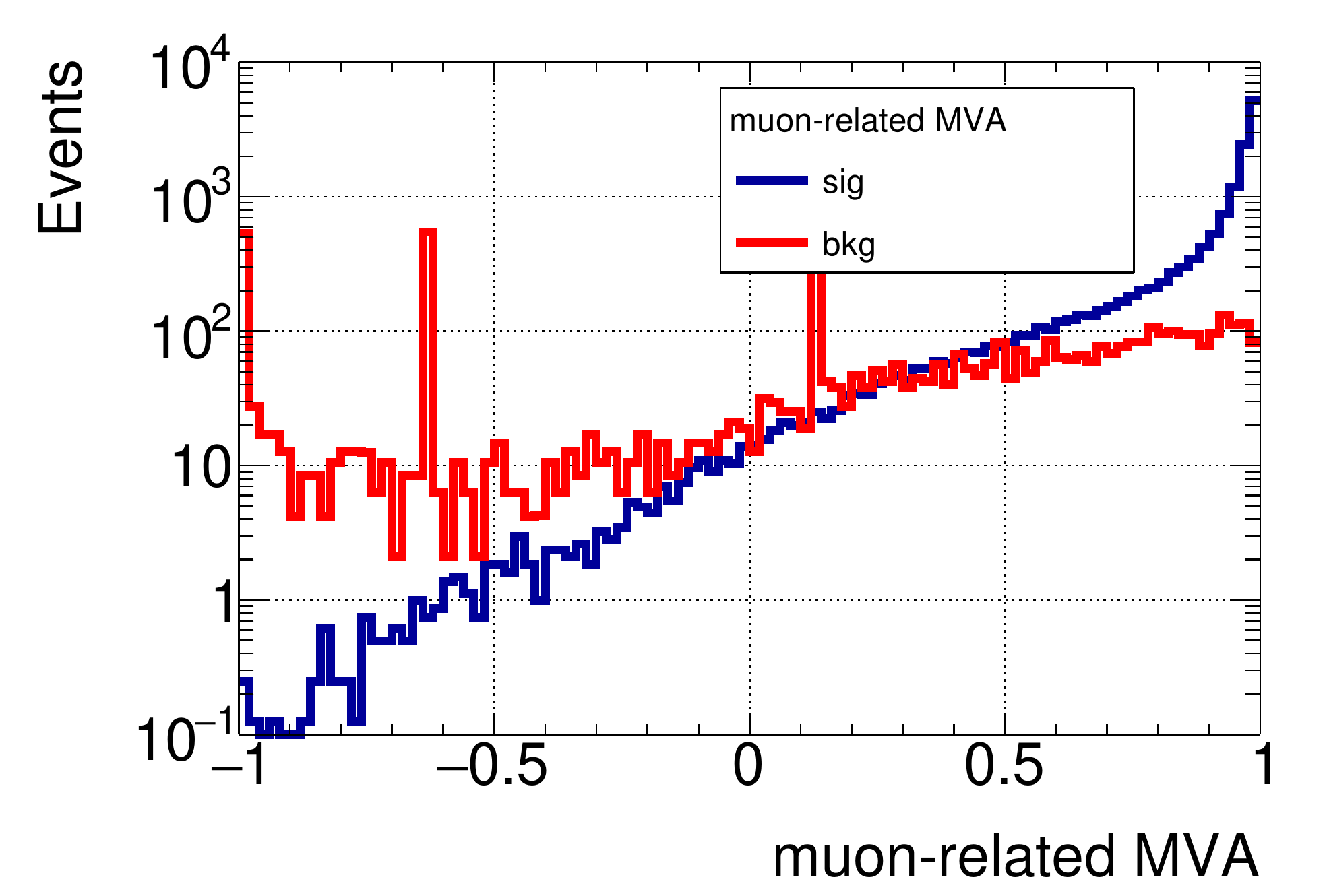}	
      \end{center}
\end{minipage}
\begin{minipage}{0.47\textwidth}
      \begin{center} 
         \includegraphics[height=5cm]{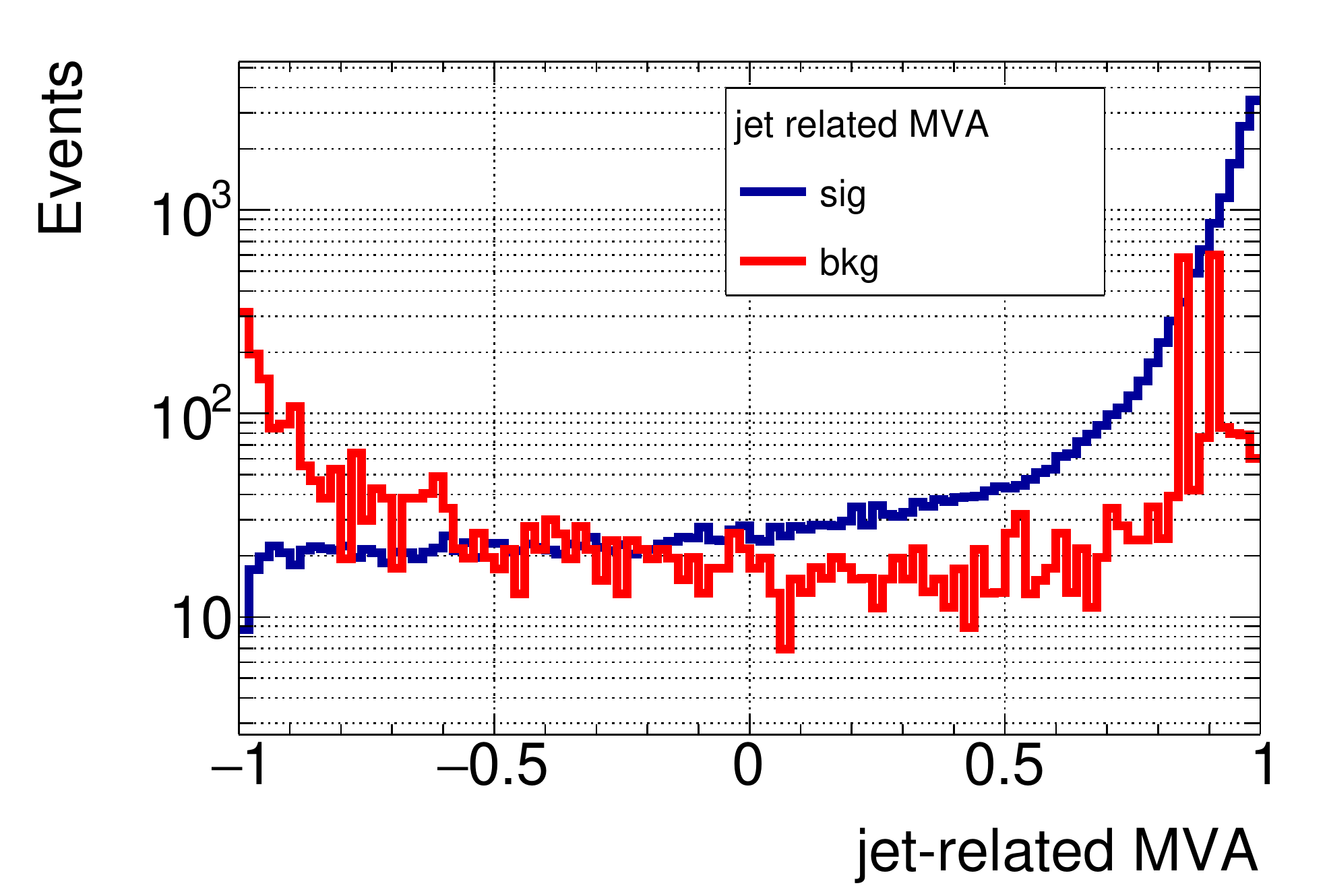}	
      \end{center}
\end{minipage}
    \caption{The lepton-related MVA output and the jet-related MVA output.}\label{f:MVA_output}
\end{figure}

The outputs of the two MVA are in Fig.~\ref{f:MVA_output}, where the signal and backgrounds are well separated.  After the kinematic cuts, the MVA cuts will be applied  to further suppress the irreducible backgrounds. 

\subsection{Event selection and results}\label{sec:simu_results}
The background suppression is performed by maximizing signal significance, which is defined as $N_{sig}/\sqrt{(N_{sig} +N_{bkg})}$, where $N_{sig}$ and $N_{bkg}$ are the event 
numbers of the signal and background processes. The
event numbers after cuts for the muon channel and electron
channel are summarized in Tables~\ref{t:table_cut_muon} and~\ref{t:table_cut_elec}, where the
luminosity is 5600 $fb^{-1}$. The significance for the $Z\to \mm $ channel is 114, while for the $Z\to \eepm $ channel is
109, which correspond to the uncertainties $\Delta\sigma(\mm h)$ =
0.88\% and $\Delta\sigma(\eepm h)$ =0.92\%. Since the $Z\to \bbbar$ is well
measured, and the Higgs decay is supposed to be the same
as the SM Higgs boson, the above uncertainty is mainly
caused by the anomalous $hZZ$ coupling. We cross-check
our results with the CEPC experimental $\mm h$ and $\eepm h$
measurements~\cite{Bai:2019qwd}. In the CEPC analysis, the Higgs
bosons decaying to$\bbbar$, $\ccbar$ and $gg$ are combined together
the significance after cuts are 96.4 for $\mm h$ and 68.3 for $\eepm h$ after extrapolating the luminosity to 5600 $fb^{-1}$.
Considering NLO effects and removing non-bjet backgrounds in their analysis, our results are consistent with
theirs.

In order to include the full detector simulation effects, we
apply the enhanced factor $k_{exp}$, which includes the detector
effects, template fit and EFT fit effects.  $k_{exp}$ is defined by $\frac{\Delta\sigma_{fast}}{\Delta\sigma_{full}}$, where $\Delta\sigma_{fast}$ is our result with the fast MC simulation
and recoil mass method, and $\Delta\sigma_{full}$ is picked from the
CEPC full simulation analysis with also the recoil mass
method~\cite{An:2018dwb}. $k_{exp}$ is about 5.3, and the detailed calculation as
well as the discussion can be found in Appendix B. After
dividing $k_{exp}$, the uncertainty of $\Delta\sigma$ with explicit final state
searching will reach 0.166\% for $\Delta\sigma(\mm h)$ and 0.173\%
for $\Delta\sigma(\eepm h)$. They are both smaller than the deviation
induced by the IDM at 240 GeV in Table~\ref{t:cross_section}. If combining the
different Higgs and $Z$ decay channels, the detective
potential of the IDM model will be further improved.
Thus, there is the opportunity to measure the deviation from
the SM model by the loop effects in the IDM model at
the CEPC.

\begin{table}
 \begin{center}
 \begin{small}
   \caption{The cut table of $Z\to\mm$ channel, when the luminosity is $  \int L dt =5600~fb^{-1}  $. }\label{t:table_cut_muon} 
 \begin{tabular}{|c| c| c| c| c| c| c| c| c| c|}
 \hline
 $                \int L dt =5600~fb^{-1}  $ &  $ \mm h_{\text{IDM}} $ &               2f  &             4f  &               Higgs  &             total backgrounds  &               efficiency &                      S/B &             significance\\ 
 \hline
  Preselection  &                    18547.4 &                     7878 &                  56776.1 &                    140.4 &                  64794.5 &                        1 &                     0.29 &                    64.25\\ 
 \hline
 $         \mmu \in [ 73 ,120 ] $ GeV&                  18000.6 &                     6060 &                  48647.6 &                    131.8 &                  54839.3 &                     0.97 &                     0.33 &                     66.7\\ 
 \hline
 $      \ptmm \in [ 10 ,70 ] $ GeV&                    17679 &                     3030 &                  38429.5 &                    129.5 &                    41589 &                     0.95 &                     0.43 &                    72.62\\ 
 \hline
 $                 E_{vis} \in [ 50 ,300 ] $ GeV&                    17679 &                     3030 &                  38429.5 &                    124.1 &                  41583.6 &                     0.95 &                     0.43 &                    72.62\\ 
 \hline
 $                \mrecmm \in [ 110 ,155 ] $ GeV&                  17665.8 &                     2424 &                   6799.5 &                      124 &                   9347.5 &                     0.95 &                     1.89 &                   107.48\\ 
 \hline
 $          \mmb \in [ 50 ,130 ] $ GeV&                  17514.9 &                     2424 &                   6306.7 &                    114.8 &                   8845.6 &                     0.94 &                     1.98 &                   107.88\\ 
 \hline
 $\mrecbb \in [ 70 ,140 ] $ GeV&                  16244.1 &                     1212 &                     4549 &                     77.5 &                   5838.5 &                     0.88 &                     2.78 &                   109.31\\ 
 \hline
 $               \text{MVA}_{\mu} \in [ -0.74 ,1 ] $ &                  16240.7 &                     1212 &                   3793.9 &                     77.5 &                   5083.3 &                     0.88 &                     3.19 &                   111.22\\ 
 \hline
 $                \text{MVA}_{j} \in [ -0.62 ,1 ] $ &                  15829.8 &                     1212 &                     2166 &                     68.1 &                   3446.1 &                     0.85 &                     4.59 &                   114.02\\ 
 \hline
 \end{tabular}
 \end{small}
 \end{center}
  \end{table}

\begin{table}
  \begin{center}
 \begin{small}
  \caption{The cut table of $Z\to\eepm$ channel, when the luminosity is $  \int L dt =5600~fb^{-1}  $. }\label{t:table_cut_elec}
 \begin{tabular}{|c| c| c| c| c| c| c| c| c|}
 \hline
 $                \int L dt =5600~fb^{-1}  $ &  $ \eepm h_{\text{IDM}} $ &               2f  &             4f  &               Higgs  &             total backgrounds  &               efficiency &                      S/B &             significance\\ 
 \hline
  Preselection                               & 18790.3 &                     9090 &                  88126.9 &                 240.6 &                  97457.5 &                        1 &                     0.19 &                    55.11\\ 
 \hline
 $         \mee \in [ 73 ,120 ] $ GeV&                  17780.2 &                     5454 &                  62034.1 &                       131 &                  67619.1 &                     0.95 &                     0.26 &                    60.84\\ 
 \hline
 $       \ptee \in [ 10 ,70 ] $ GeV&                  17439.3 &                     2424 &                  51180.8 &                   128.7 &                  53733.5 &                     0.93 &                     0.32 &                    65.37\\ 
 \hline
 $                 E_{vis} \in [ 50 ,300 ] $ GeV&                  17439.3 &                     2424 &                  51176.5 &                     123.6 &                  53724.1 &                     0.93 &                     0.32 &                    65.37\\ 
 \hline
 $               \mrecee \in [ 110 ,155 ] $ GeV&                  17411.8 &                      606 &                   8772.4 &                      123.4 &                   9501.8 &                     0.93 &                     1.83 &                   106.13\\ 
 \hline
 $           \mmb\in [ 50 ,130 ] $ GeV&                  17183.7 &                      606 &                   8218.9 &                     114.4 &                   8939.3 &                     0.91 &                     1.92 &                   106.32\\ 
 \hline
 $\mrecbb \in [ 70 ,140 ] $ GeV&                  15960.8 &                      606 &                   6151.0 &                     76.6 &                   6833.6 &                     0.85 &                     2.34 &                   105.72\\ 
 \hline
 $               \text{MVA}_{e} \in [ -0.74 ,1 ] $ &                  15959.1 &                      606 &                   5060.1 &                       76.6 &                   5742.7 &                     0.85 &                     2.78 &                   108.33\\ 
 \hline
 $                \text{MVA}_{j} \in [ -0.62 ,1 ] $ &                    15905 &                      606 &                   4604.6 &                      75.4 &                     5286 &                     0.85 &                     3.01 &                   109.26\\ 
 \hline
  \end{tabular}
 \end{small}
 \end{center}
  \end{table}

\section{Conclusion}\label{sum}
We have performed the MC simulation of the lepton
collider signals at electroweak one-loop level at future
lepton colliders in synergy with the GW signals. The
signals at future GW detectors and lepton colliders could
make complementary exploration on the blind spots of this
DM model. There is the opportunity to measure the
deviation from the SM model by the loop effects in the
IDM model at the future lepton colliders. In the future, if
we observe the predicted GW signal at U-DECIGO, we
would expect that the corresponding collider signals could
be observed at the future lepton collider, and vice versa.
Based on the study here, we will investigate more generic
DM models with the blind spots, which might give more
stronger collider and GW signals.

\section*{Acknowledgements}
We would like to thank Manqi Ruan for valuable discussions on the performance of CEPC project.
 Y. W. is supported by the ‘Scientific Research Funding Project for Introduced High-level Talents’ of the
Inner Mongolia Normal University Grant No. 2019YJRC001,  and the scientific research
funding for introduced high-level talents of Inner Mongolia of China.
C.S.L. is supported by the National Nature Science foundation of China, under Grants No. 11875072.
F.P.H. is supported in part by the initial funding of 
Sun Yat-Sen University,   Guangdong Major Project of Basic and Applied Basic Research (Grant No. 2019B030302001), and the McDonnell Center for the Space Sciences.
 
\appendix

\section{Strong first-order phase transition}
To discuss the phase transition dynamics in the IDM,
we first write the Higgs doublet field $\Phi$ in terms of the
background field $h$, namely,
\be
\Phi=	\begin{pmatrix}
 		0\\
 		\frac{1}{\sqrt{2}}h  
	\end{pmatrix} \,\,.
\ee
Further, the effective potential at the finite temperature can
be obtained as
\begin{equation}
V_{\mathrm{eff}}(h,T) = V_{0}(h)+V_{\rm CW}(h) +
V_{\rm ther}(h, T) +V_{\rm daisy}(h, T) \,\,.\nonumber
 \label{veff}
\end{equation}
$V_0(h)=\frac{\mu_1^2 h^2}{2}+\frac{\lambda_1 h^4}{4}$ is the tree-level potential.
$V_{\rm CW}(h)$ is the Coleman-Weinberg potential at zero temperature.
$V_{\rm ther}(h, T)$ is the thermal correction.
$V_{\rm daisy}(h, T)$
represents the daisy resummation.The state-of-the-art calculations of
the finite-temperature effective potential and its phase
transition behavior are the recent two-loop investigations
by Refs.~\cite{Laine:2017hdk,Senaha:2018xek}. Their results show that the one-loop
effective potential in the high temperature expansion is
rather reliable in the IDM~\cite{Laine:2017hdk,Senaha:2018xek} and the corrections
compared to one-loop results are small~\cite{Laine:2017hdk,Senaha:2018xek}. To clearly
see the phase transition dynamics and simplify the following discussions on the phase transition GW signals, we only
consider the one-loop effective potential including the
daisy resummation. Since we only consider the IDM, only
the Higgs doublet gets VEV. The phase transition along the
Higgs field direction is favored in the benchmark points,
which is well studied in previous literature~\cite{Chowdhury:2011ga,Borah:2012pu,
Gil:2012ya,Cline:2013bln,AbdusSalam:2013eya,
Blinov:2015vma,Cao:2017oez,Huang:2017rzf,
Laine:2017hdk,Senaha:2018xek,Huang:2019riv,
Kainulainen:2019kyp}.

The leading-order thermal corrections to the effective potential in the Landau gauge can be written as
\beq
V_{\rm ther}(h, T) =
\frac{T^4}{2\pi^2}\left(\sum_{i=\mathrm{bosons}} n_i J_B\left[m^2_i(h)/T^2\right]
+
\sum_{i=\mathrm{fermions}} n_i J_F\left[m^2_i(h)/T^2\right]
\right),
\eeq
where the $J$ functions are defined as
\begin{align}
J_B(x) & = \int_0^\infty dt\; t^2 \ln\left[1 - \exp\left(-\sqrt{t^2 + x}\right)\right]\label{eq:JB}\,\,,\\
J_F(x) & = \int_0^\infty dt\; t^2 \ln\left[1 + \exp\left(-\sqrt{t^2 + x}\right)\right]\,\,.
\end{align}
Under high-temperature expansions, we have
\begin{align}
T^4 J_B\left[m^2/T^2\right] & = -\frac{\pi^4 T^4}{45}+
\frac{\pi^2}{12} T^2 m^2-\frac{\pi}{6}
T \left(m^2\right)^{3/2}-\frac{1}{32}
m^4\ln\frac{m^2}{a_b T^2} + \mathcal{O}\left(m^2/T^2\right)\label{eq:Jb_highT}\,\,,\\
T^4 J_F\left[m^2/T^2\right]& = \frac{7\pi^4 T^4}{360}-
\frac{\pi^2}{24} T^2 m^2-\frac{1}{32}
m^4\ln\frac{m^2}{a_f T^2} + \mathcal{O}\left(m^2/T^2\right)\label{eq:Jf_highT}\,\,,
\end{align}
where $a_b = 16 a_f = 16\pi^2 \exp(3/2 - 2\gamma_E)$. 
In the above
definition, the degree of freedom for the fermions is a
negative integer to ensure a positive $T^2$ term. The positive
$T^2$ terms for both bosons and fermions in the above
expressions enable the symmetry restoration at high temperatures. The nonanalytic $m^3$ term in Eq.~\eqref{eq:Jb_highT} can be
responsible for the thermal barrier and the SFOPT between
the high-temperature phase and the low-temperature phase.

The field-dependent masses of the gauge bosons and the
top quark at zero temperature are given by

\bea
m_{W}^2(h) &=& \frac{g^2}{4} h^2, \
m_{Z}^2(h) = \frac{g^2+g'^2}{4} h^2,
m_{t}^2(h) = \frac{y_t^2}{2} h^2, \nn
\label{masses}
\eea
where $y_t$ is the top Yukawa coupling, $g$ and $g'$ are the gauge coupling of  $SU(2)_L$ and  $U(1)_Y$ gauge group, respectively.
The field-dependent thermal scalar masses are
\begin{align}
	m_h^2&=\lambda_1 h^2 \label{eq:mhsq}\,\,,\\
	m_H^2&=\mu_2^2+\frac{1}{2}(\lambda_3+\lambda_4+\lambda_5)h^2\,\,, \\
	m_A^2&=\mu_2^2+\frac{1}{2}(\lambda_3+\lambda_4-\lambda_5)h^2\,\,, \\
   m^2_{H^+}&=\mu_2^2+\frac{1}{2}\lambda_3 h^2\,\,.
\end{align}

In the above formulas, we have considered the contribution from daisy resummation in the Arnold-Espinosa
scheme, which reads as
\bea
	V_{daisy} & \supset & - \frac{T}{12\pi} \sum_{i = {\rm b}} n_b\left(\left[m_i^{2}(h,T)\right]^{3/2} -  \left[m_i^{2}(h)\right]^{3/2} \right) \,\,. \nonumber
\eea
Here, the thermal field-dependent masses
$m_i^2(h, T) \equiv m_i^2(h)+ \Pi_i(h, T)$,
where $\Pi_i (h, T)$ is the bosonic field $i$'s self-energy in the IR limit.

All the scalar particles can get a thermal mass by replacing
\begin{equation}
\mu_2^2 \to \mu_2^2+c_2 T^2
\end{equation}
\begin{equation}
\mu_1^2 \to \mu_1^2+c_1 T^2
\end{equation}
with the thermal correction coefficients
\begin{equation}
c_1 = \frac{3\lambda_1+2\lambda_{3}+\lambda_4}{12} + \frac{3g^2 + g^{'2}}{16}
	+ \frac{y_t^2}{4}\,\,,
\end{equation}
\begin{equation}
c_2=  \frac{3\lambda_{2}+2\lambda_{3}+\lambda_4}{12} + \frac{3g^2 + g^{'2}}{16}\,\,.
\end{equation}

There is no thermal mass corrections for the fermions and the transverse component of the gauge bosons at leading order.
Only the longitudinal component of the gauge boson has the thermal mass corrections, namely,
\begin{equation}
\Pi_W(T)=2 g^2 T^2\,\,,
\end{equation}
\begin{equation}
\Pi_B(T)=2 g^{'2} T^2\,\,.
\end{equation}

To calculate the effective potential in IDM, the degrees of freedom for each particles running in the loop are shown below:
\bea
n_{W^{\pm}}=4, \  n_{Z}=2, \  n_{\pi}=3\,\,, \nonumber \\
\  n_h=n_{H} = n_{H^+}=n_{H^+} = 1, \  n_{t}=-12\,\,. \nonumber
\eea

\section{The analysis with the model-independent method}\label{app:recoil}
We perform the model-independent measurement for the
 $hZZ$ coupling in Table~\ref{t:table_cut_muon_recoil}. The analysis algorithms are
the same as the CEPC/ILC experimental analysis with the
recoil mass technique. In the whole analysis, only $Z$ decay
products, i.e., $\mm$, are used to constitute the kinematics as
$\mmu$, $\ptmm$, no information from the h decay products are
involved. Thus, the measurement for $hZZ$ coupling would
not depend on model-specific assumptions on the properties of the Higgs boson. However, this method will also
decrease the significance when searching new phenomena.
For example, in the following analysis, with the same
datasets, the maximum significance is about 38, which is
obviously smaller than the results with explicit Higgs decay
final states in Table~\ref{t:table_cut_muon}.

 The corresponding uncertainty with model-independent
measurement is $\Delta\sigma\sim 2.6\%$. It is similar to the result in the
CEPC report~\cite{CEPCStudyGroup:2018ghi}, which only applied the kinematic cuts. In
our paper, the parton-level event generation and hadronizations are completely the same with the generation in
CEPC, including the luminosity, the simulation event
number and weight, and so on. The difference only comes
from the detector simulation and following signal analysis.
In CEPC collaboration’s analysis with full detector simulation events, after the normal kinematic cuts, they applied
the template fit and the parameter fit to combine the
CEPC and HL-LHC data~\cite{An:2018dwb}, then $\Delta\sigma$ will further reach
the report value 0.5\%. In this work, we use the CEPC
default Delphes Card for detector simulation in Delphes and
we only add normal kinematic cuts for signal analysis,
without template fit and the parameter fit. As a result,
comparing our results and the CEPC results, the ratio of the
significance would reflect the differences from the detector
simulation and the following analysis procedures. Thus, in
this note, we define the enhanced factor $k_{exp}$ to include the
full-simulation effects, template fit cut effect, EFT fitting
with LHC data and all other related effects. $k_{exp}$ of the
CEPC collaboration reported $\Delta\sigma$ to our simulated value is
about 5.3. We also assume $k_{exp}$ is the same for the $Z\to \eepm$ channel and the $Z\to \mm$ channel. We will use this
enhanced factor to estimate the measurement uncertainties
of $hZZ$ with the explicit Higgs decay final states at the
CEPC in Sec.~\ref{sec:simu_results}.

 \begin{table}
\begin{center}
 \begin{small}
  \caption{The cut table of $Z\to\mm$ channel with a model-independent measurements, when the luminosity is $  \int L dt =5600~fb^{-1}  $. }\label{t:table_cut_muon_recoil}
 \begin{tabular}{|c| c| c| c| c| c| c| c| c|}
 \hline
 $                \int L dt =5600~fb^{-1}  $ &  $ \mm h_{\text{IDM}} $ &               2f  &             4f  &               Higgs  &             total  &               efficiency &                      S/B &             significance\\ 
 \hline
 Preselection &                    19841 &              1.89$\times 10^{6}$ &                   717156 &                  12195.9 &              2.62$\times 10^{6}$ &                        1 &                    0.008 &                    12.21\\ 
 \hline
 $         \mmu \in [ 73 ,120 ] $ GeV&                  19238.5 &              1.51$\times 10^{6}$ &                   486730 &                  11764.5 &              2.01$\times 10^{6}$ &                     0.97 &                     0.01 &                    13.52\\ 
 \hline
 $       \ptmm \in [ 10 ,70 ] $ GeV&                  18890.9 &                   481428 &                   402934 &                  11552.8 &                   895915 &                     0.95 &                     0.02 &                    19.75\\ 
 \hline
 $                 E_{vis} \in [ 50 ,300 ] $ GeV&                  18890.9 &                   466789 &                   308576 &                  11187.9 &                   786553 &                     0.95 &                     0.02 &                    21.05\\ 
 \hline
 $               \text{MVA}_{\mu} \in [ -0.74 ,1 ] $ &                  18877.7 &                   158664 &                  64050.7 &                  10582.6 &                   233297 &                     0.95 &                     0.08 &                    37.59\\ 
 \hline
 $           \mrecmm \in [ 110 ,155 ] $ GeV&                  18864.4 &                   154942 &                    62779 &                  10575.7 &                   228297 &                     0.95 &                     0.08 &                    37.94\\ 
 \hline
 \end{tabular}
 \end{small}
 \end{center}
  \end{table}

\bibliography{references}
\end{document}